# High accuracy, absolute, cryogenic refractive index measurements of infrared lens materials for JWST NIRCam using CHARMS

Douglas B. Leviton[a*], Bradley J. Frey[a], Todd Kvamme[b]
[a]NASA Goddard Space Flight Center, Greenbelt, MD  20771
[b]Lockheed Martin Corporation, Palo Alto, CA  94304

## ABSTRACT

The refractive optical design of the James Webb Space Telescope (JWST) Near Infrared Camera (NIRCam) uses three infrared materials in its lenses: LiF, BaF$_2$, and ZnSe.  In order to provide the instrument's optical designers with accurate, heretofore unavailable data for absolute refractive index based on actual cryogenic measurements, two prismatic samples of each material were measured using the cryogenic, high accuracy, refraction measuring system (CHARMS) at NASA's Goddard Space Flight Center (GSFC), densely covering the temperature range from 15 to 320 K and wavelength range from 0.4 to 5.6 microns.  Data reduction methods are discussed and graphical and tabulated data for absolute refractive index, dispersion, and thermo-optic coefficient for these three materials are presented for selected wavelengths and temperatures along with estimates of index uncertainty.   Coefficients for temperature-dependent Sellmeier fits of measured index are also presented with an example of their usage to predict absolute index at any wavelength or temperature within the applicable range of those parameters.

**Keywords:** Cryogenic, refractive index, infrared, refractometer, NIRCam, CHARMS, Sellmeier, lithium fluoride, barium fluoride, zinc selenide

## 1. INTRODUCTION

The Near Infrared Camera (NIRCam) is not only the primary science camera for the James Webb Space Telescope (JWST), it is also the observatory's wavefront sensor.  The camera is designed to function at a nominal operating temperature of 37 K.  It contains two essentially identical imaging channels which are fed by a dichroic beamsplitter, which separates incoming light from the JWST optical telescope into two wavelength bands from 0.6 to 2.3 μm and from 2.3 to 5 μm.  Each channel contains two triplet lenses (a collimating lens group and a camera lens group) consisting of three different materials – lithium fluoride (LiF), barium fluoride (BaF$_2$), and zinc selenide (ZnSe) – which re-image the focal surface of the telescope through spectral bandpass filters to the channel's HgCdTe image sensor.  Each channel's design wavefront error is specified to be less than or equal to 69 nm rms.  This places stringent requirements on knowledge of the absolute refractive indices of the camera's lens materials at the operating temperature.  The requirement on knowledge of refractive index at 37 K for each material is of the order of +/-0.0001.

Historically, few accurate refractive index measurements of infrared materials have been made at cryogenic temperatures.  This has hampered the developments of many cryogenic infrared instruments, including the Infrared Array Camera (IRAC) for NASA's Spitzer Space Telescope, for which, for design purposes, index values for its lens materials were extrapolated from literature values both in wavelength and in temperature.  Such an approach leads to integration cycles which are far longer than anticipated, where best instrument performance is achieved by trial and error in many time-consuming and expensive iterations of instrument optical alignment.

The Cryogenic High Accuracy Refraction Measuring System (CHARMS) has been recently developed at GSFC to make such measurements in the absolute sense (in vacuum) down to temperatures as low as 15 K with unsurpassed accuracy using the method of minimum deviation refractometry.[1,2,3]  For low index materials with only modest dispersion such as synthetic fused silica, CHARMS can measure absolute index with an uncertainty in the sixth decimal place of index.  For materials with higher indices and high dispersion such as CVD ZnSe, CHARMS can measure absolute index with an uncertainty of about 1 part in the fourth decimal place of index.  Measurement methods used and recent facility improvements are discussed in a companion paper.[4]

---

* doug.leviton@nasa.gov, phone 1-301-286-3670, FAX 1-301-286-0204

## 2. TEST SPECIMENS

Prism pairs purchased for this study were fabricated from one boule of each respective material. Therefore, this study is not primarily an interspecimen variability study for specimens from different material suppliers or from different lots of material from a given supplier. Study of such variability is planned in future efforts.

One prism of each material is shown in the photo in Figure 1. The apex angle of the prism for each material is designed so that the beam deviation angle for the highest index in the material's transparent range will equal the largest accessible deviation angle of the refractometer, 60°. In the figure, the prisms are illuminated with collimated white light. The ZnSe prism on the left, whose wavelength cutoff is in the mid-visible, can be seen to deviate light through a much larger angle than do the LiF and $BaF_2$ prisms whose wavelength cutoffs are deep in the far ultraviolet, where the refractometer will have spectral coverage in the future. Common prism dimensions are refracting face length and height of 38.1 mm and 28.6 mm, respectively. Nominal apex angles of these prisms are: ZnSe – 29.0°; $BaF_2$ – 52.0°; and LiF – 60.0°.

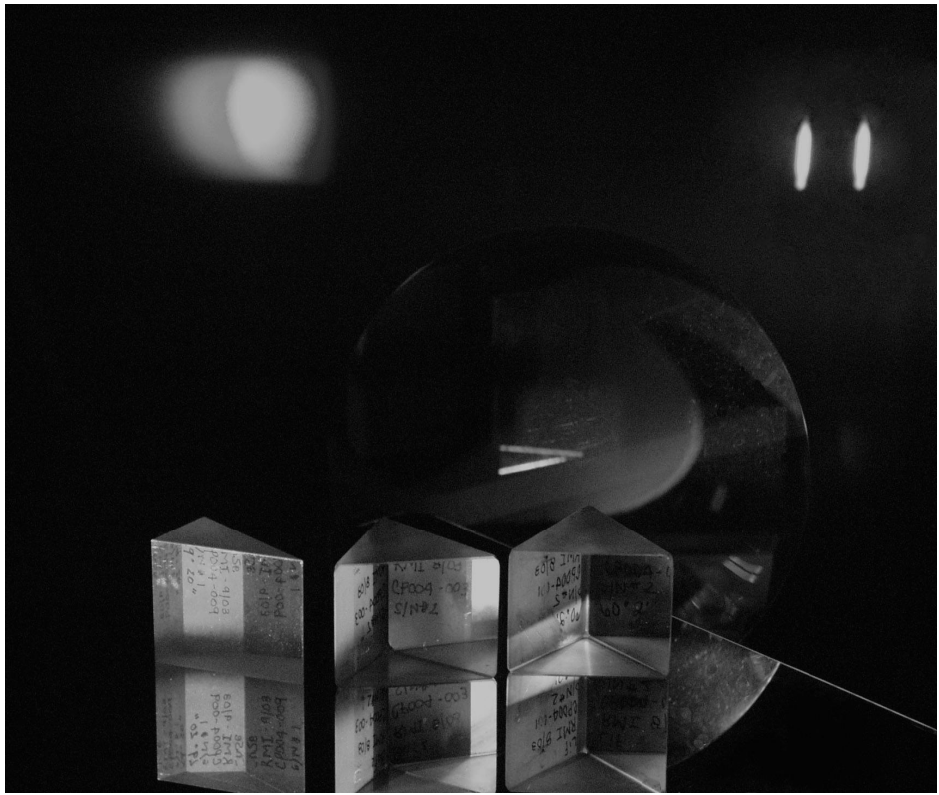

Figure 1 – Representative prismatic specimens studied: (L-to-R: ZnSe, $BaF_2$, and LiF). The prisms (which rest on the surface of a flat mirror) are illuminated near minimum deviation with a collimated beam of white light coming from the lower left corner of the photo. The dispersion of ZnSe in the visible is high enough that a clearly color-gradated spectrum can be viewed on a screen – even without a re-imaging lens – down to the material's short wavelength cutoff at about 0.53 microns (blue-green) (upper left corner of photo). The beams through the $BaF_2$ and LiF prisms pass through a large lens which forms re-imaged, full spectra of the light source on the screen (upper right corner of photo). The ground, non-refracting face of each prism is labeled with identifying information.

## 3. REFRACTOMETER CALIBRATION AND TREATMENT OF MEASURED INDEX DATA

By way of refractometer calibration, it was incumbent on us to at least demonstrate index accuracy at room temperature to the level of our estimated uncertainty for a material for which some trusted, pre-existing truth was available. To that end, we measured the absolute refractive index of synthetic fused silica and compared those results to the well-known dispersion law of Malitson[5]. Our estimated uncertainty for fused silica is about $1\text{-}1.5 \times 10^{-5}$ at room temperature. After adjusting Malitson's values by the index of air (in which his measurements were performed), our measured indices typically agreed with Malitson's dispersion law to within $8 \times 10^{-6}$ over the wavelength range of applicability of that law.

We use a computer program we call the CHARMS Data Cruncher (CDC) to examine raw data from the refractometer and reduce it to the point where resulting measured index values can be fit to Sellmeier equations. During CDC runs, measured index values for each wavelength are fit piecewise to second order polynomials in temperature above and below some selected crossover temperature, T, to get in-process assessments of data quality. These high quality quadratic fits are of the form $n(T) = c_2 T^2 + c_1 T + c_0$ and apply to temperatures above what we call the saturation temperature which is that temperature below which we can no longer definitively sense a change in index. We call the saturation value of index that index which we measure at and below the saturation temperature. All materials we have studied exhibit a saturation temperature, except for possibly ZnSe. To date, our determination of the saturation temperatures for $BaF_2$, LiF, ZnSe, and fused silica are about 50 K, 65 K, 20 K, and 32 K, respectively.

In order to compute spectral dispersion, first a table of index values is computed on a regular wavelength/temperature grid from the piecewise quadratic fits described above. From that table, a new table of spectral dispersion, $dn/d\lambda$, is computed by dividing differences in index value, n, by corresponding differences in wavelength, $\lambda$, for each temperature. Thermo-optic coefficient, $dn/dT$, is simply the first derivative of n(T) with respect to T or $dn(T)/dT = 2 c_2 T + c_1$. CDC produces a table of thermo-optic coefficients on the same regular wavelength and temperature grid described above.

Finally, CDC produces a table of estimated index errors for different wavelength and temperature combinations. A partial index error dn is computed for each of four factors (based on presumably known uncertainties in those factors), and the four resulting dn's are combined in quadrature to produce a net index error estimate. The four partial dn's are computed using: 1) uncertainty in calibrated wavelength, $d\lambda$, along with computed $dn/d\lambda$; 2) uncertainty in measured temperature, dT, along with computed $dn/dT$; 3) uncertainty in measured apex angle, $d\alpha$, along with analytically derived $dn/d\alpha$; and 4) uncertainty in measured beam deviation angle, $d\delta$, along with analytically derived $dn/d\delta$.

Subsequently, raw measured indices for each material are fit to a Sellmeier model of the form:

$$n^2(\lambda,T) - 1 = \sum_{i=1}^{m} \frac{S_i(T) \cdot \lambda^2}{\lambda^2 - \lambda_i^2(T)}$$

where $S_i$ are the strengths of the resonance features in the material at wavelengths $\lambda_i$. When dealing with a wavelength interval between wavelengths of physical resonances in the material, the summation may be approximated by only a few terms, m – typically three.[6] In such an approximation, resonance strengths $S_i$ and wavelengths $\lambda_i$ no longer have direct physical significance but are rather parameters used to generate an adequately accurate fit to empirical data. If these parameters are assumed to be functions of T, one can generate a temperature-dependent Sellmeier model for $n(\lambda,T)$.

Historically, this modeling approach has been employed with significant success for a variety of materials despite a rather serious sparsity of available index measurements – to cover a wide range of temperatures and wavelengths – upon which to base a model. One solution to the shortcoming of lack of measured index data has been to appeal to room temperature refractive index data at several wavelengths to anchor the model and then to extrapolate index values for other temperatures using accurate measurements of the thermo-optic coefficient dn/dT at those temperatures, which are much easier to make than accurate measurements of the index itself at exotic temperatures.[6] So, for purposes of NIRCam's lens designs, historical index models, which are based on dn/dT measurements made at temperatures only as low as 93 K, are inherently limited in applicability at NIRCam's operating temperature of 37 K. Indeed, those models do not predict the existence of a relevant saturation index which at least two of NIRCam's lens materials possess.

Meanwhile, with CHARMS, we have made direct measurements of index densely sampled over a wide range of wavelengths and temperatures to produce a model with residuals on the order of the uncertainties in our raw index measurements. For our models, we have found that 4th order temperature dependences in all three terms in each of $S_i$ and $\lambda_i$ work adequately well, as also found previously in the literature. The Sellmeier equation consequently becomes:

$$n^2(\lambda,T) - 1 = \sum_{i=1}^{3} \frac{S_i(T) \cdot \lambda^2}{\lambda^2 - \lambda_i^2(T)}$$

where,

$$S_i(T) = \sum_{j=0}^{4} S_{ij} \cdot T^j$$

$$\lambda_i(T) = \sum_{j=0}^{4} \lambda_{ij} \cdot T^j$$

As an example of how to apply the Sellmeier coefficients to compute index of a material at a particular wavelength and temperature, suppose we desire to obtain the absolute refractive index of $BaF_2$ at a wavelength of 1.0 μm and at a temperature of 77 K.

$$S_i(T) = \sum_{j=0}^{4} S_{ij} \cdot T^j = S_{i0} + S_{i1} \cdot T + S_{i2} \cdot T^2 + S_{i3} \cdot T^3 + S_{i4} \cdot T^4$$

Using the coefficients in Table 5 for $BaF_2$, the reader can to verify that:

$S_1(77\ K) = 0.7661482 + 0.000214836 \cdot (77) + 7.31813E\text{-}07 \cdot (77)^2 - 1.569E\text{-}09 \cdot (77)^3 - 2.67277E\text{-}11 \cdot (77)^4$
$\phantom{S_1(77\ K)} = 0.785374$

Similarly, $S_2(77\ K) = 0.374300$, $S_3(77\ K) = 3.89192$, $\lambda_1(77\ K) = 0.0973861$, $\lambda_2(77\ K) = 0.0466454$, $\lambda_3(77\ K) = 46.5379$. Expanding the Sellmeier summation into its three terms, we have:

$$n^2(\lambda, T) - 1 = \frac{S_1(T) \cdot \lambda^2}{\lambda^2 - \lambda_1^2(T)} + \frac{S_2(T) \cdot \lambda^2}{\lambda^2 - \lambda_2^2(T)} + \frac{S_3(T) \cdot \lambda^2}{\lambda^2 - \lambda_3^2(T)}$$

and so,

$$n(1.0\mu m, 77K) = \sqrt{1 + \frac{0.785374 \cdot (1.0)^2}{(1.0)^2 - (0.0973861)^2} + \frac{0.374300 \cdot (1.0)^2}{(1.0)^2 - (0.0466454)^2} + \frac{3.89192 \cdot (1.0)^2}{(1.0)^2 - (46.5379)^2}} = 1.471806$$

These Sellmeier models are our best statistical representation of the measured data over the complete measured ranges of wavelength and temperature. Depending on the sample material measured, the residuals of the measured values compared to the fits can be as low as several parts in the sixth decimal place of index. This level of fit quality for data obtained using CHARMS generally pertains to low index materials with moderately low dispersion such as fused silica, LiF, and $BaF_2$. For high index materials with high dispersion such as ZnSe, residuals are higher at a few parts in the fifth decimal place of index. Table 1 lists a handful of calculated values for the three NIRCam lens materials.

Table 1 – absolute refractive indices for NIRCam lens materials for selected wavelengths and temperatures computed from Sellmeier models

| **BaF2** | 50 K* | 150 K | 295 K |
| --- | --- | --- | --- |
| 0.6 microns | 1.47743 | 1.47669 | 1.47456 |
| 1.0 microns | 1.47191 | 1.47114 | 1.46901 |
| 5.0 microns | 1.45425 | 1.45350 | 1.45147 |
| **LiF** | 65 K* | 150 K | 295 K |
| 0.6 microns | 1.39468 | 1.39423 | 1.39221 |
| 1.0 microns | 1.39001 | 1.38955 | 1.38752 |
| 5.0 microns | 1.32851 | 1.32817 | 1.32680 |
| **ZnSe** | 50 K | 150 K | 295 K |
| 0.6 microns | 2.59146 | 2.59929 | 2.61488 |
| 1.0 microns | 2.47433 | 2.47976 | 2.48991 |
| 5.0 microns | 2.41709 | 2.42175 | 2.43023 |

* saturation temperature for this material

## 4. REFRACTIVE INDICES FOR NIRCAM LENS MATERIALS

### 4.1 Barium Fluoride (BaF$_2$)

Absolute refractive indices of BaF$_2$ were measured over the 0.45 to 5.6 microns wavelength range and over the temperature range from 21 to 300 K for two test specimens which yielded equal indices to within our measurement uncertainty of +/-1.5 x 10$^{-5}$. Indices are tabulated in Table 2, plotted in Figure 2 for selected temperatures and wavelengths. Spectral dispersion is tabulated in Table 3, plotted in Figure 3. Thermo-optic coefficient is tabulated in Table 4, plotted in Figure 4. Coefficients for the three term Sellmeier model with 4$^{th}$ order temperature dependence are given in Table 5. We compared our Sellmeier model of BaF$_2$ at room temperature to the index measured by Malitson[7] in air multiplied by the index of air (assumed to be a constant equal to 1.00027). Our model agrees with Malitson's model at room temperature to within ~1 x 10$^{-5}$ at all wavelengths. This is actually somewhat astounding in that those measurements were performed on a sample of BaF$_2$ made at least 40 years ago.

Table 2 – absolute refractive index (n) for BaF$_2$ at selected wavelengths and temperatures

| wavelength | 50 K | 60 K | 70 K | 80 K | 90 K | 100 K | 110 K | 120 K | 150 K | 200 K | 250 K | 275 K | 295 K | 300 K |
|---|---|---|---|---|---|---|---|---|---|---|---|---|---|---|
| 0.45 microns | 1.48401 | 1.48398 | 1.48394 | 1.48389 | 1.48383 | 1.48376 | 1.48368 | 1.48359 | 1.48325 | 1.48260 | 1.48187 | 1.48149 | 1.48116 | 1.48108 |
| 0.50 microns | 1.48114 | 1.48111 | 1.48107 | 1.48102 | 1.48096 | 1.48089 | 1.48081 | 1.48071 | 1.48037 | 1.47971 | 1.47898 | 1.47860 | 1.47827 | 1.47819 |
| 0.60 microns | 1.47744 | 1.47740 | 1.47737 | 1.47732 | 1.47726 | 1.47719 | 1.47711 | 1.47701 | 1.47667 | 1.47600 | 1.47527 | 1.47488 | 1.47456 | 1.47448 |
| 0.70 microns | 1.47521 | 1.47518 | 1.47514 | 1.47510 | 1.47504 | 1.47497 | 1.47488 | 1.47479 | 1.47444 | 1.47377 | 1.47303 | 1.47264 | 1.47232 | 1.47223 |
| 0.80 microns | 1.47374 | 1.47371 | 1.47367 | 1.47362 | 1.47356 | 1.47349 | 1.47341 | 1.47331 | 1.47296 | 1.47229 | 1.47156 | 1.47116 | 1.47084 | 1.47075 |
| 0.90 microns | 1.47270 | 1.47266 | 1.47263 | 1.47258 | 1.47252 | 1.47245 | 1.47236 | 1.47227 | 1.47192 | 1.47125 | 1.47051 | 1.47012 | 1.46979 | 1.46971 |
| 1.00 microns | 1.47191 | 1.47188 | 1.47184 | 1.47179 | 1.47173 | 1.47166 | 1.47158 | 1.47148 | 1.47113 | 1.47046 | 1.46972 | 1.46933 | 1.46900 | 1.46892 |
| 1.20 microns | 1.47079 | 1.47074 | 1.47070 | 1.47065 | 1.47058 | 1.47051 | 1.47042 | 1.47033 | 1.46999 | 1.46933 | 1.46860 | 1.46820 | 1.46788 | 1.46779 |
| 1.50 microns | 1.46959 | 1.46954 | 1.46949 | 1.46944 | 1.46938 | 1.46930 | 1.46922 | 1.46912 | 1.46878 | 1.46812 | 1.46739 | 1.46700 | 1.46667 | 1.46658 |
| 2.00 microns | 1.46796 | 1.46791 | 1.46787 | 1.46782 | 1.46775 | 1.46768 | 1.46759 | 1.46750 | 1.46716 | 1.46650 | 1.46577 | 1.46538 | 1.46506 | 1.46497 |
| 2.50 microns | 1.46632 | 1.46628 | 1.46623 | 1.46618 | 1.46611 | 1.46604 | 1.46595 | 1.46586 | 1.46553 | 1.46487 | 1.46414 | 1.46375 | 1.46342 | 1.46334 |
| 3.00 microns | 1.46449 | 1.46443 | 1.46439 | 1.46434 | 1.46428 | 1.46420 | 1.46412 | 1.46402 | 1.46369 | 1.46303 | 1.46231 | 1.46192 | 1.46159 | 1.46151 |
| 3.50 microns | 1.46238 | 1.46234 | 1.46230 | 1.46224 | 1.46218 | 1.46211 | 1.46202 | 1.46193 | 1.46159 | 1.46094 | 1.46022 | 1.45983 | 1.45951 | 1.45943 |
| 4.00 microns | 1.45999 | 1.45995 | 1.45991 | 1.45985 | 1.45979 | 1.45972 | 1.45963 | 1.45954 | 1.45921 | 1.45856 | 1.45784 | 1.45745 | 1.45714 | 1.45705 |
| 4.50 microns | 1.45729 | 1.45725 | 1.45720 | 1.45715 | 1.45709 | 1.45702 | 1.45693 | 1.45684 | 1.45651 | 1.45586 | 1.45515 | 1.45477 | 1.45446 | 1.45438 |
| 5.00 microns | 1.45427 | 1.45423 | 1.45418 | 1.45413 | 1.45407 | 1.45400 | 1.45392 | 1.45383 | 1.45350 | 1.45286 | 1.45215 | 1.45178 | 1.45147 | 1.45139 |
| 5.50 microns | 1.45091 | 1.45087 | 1.45083 | 1.45078 | 1.45072 | 1.45065 | 1.45057 | 1.45048 | 1.45015 | 1.44951 | 1.44882 | 1.44845 | 1.44814 | 1.44806 |

Table 3 – spectral dispersion (dn/dλ) in BaF$_2$ at selected wavelengths and temperatures in units of 1/microns

| wavelength | 50 K | 60 K | 70 K | 80 K | 90 K | 100 K | 110 K | 120 K | 150 K | 200 K | 250 K | 275 K | 295 K | 300 K |
|---|---|---|---|---|---|---|---|---|---|---|---|---|---|---|
| 0.50 microns | -0.0497 | -0.0498 | -0.0498 | -0.0498 | -0.0498 | -0.0498 | -0.0498 | -0.0498 | -0.0497 | -0.0499 | -0.0500 | -0.0501 | -0.0501 | -0.0501 |
| 0.60 microns | -0.0284 | -0.0284 | -0.0283 | -0.0283 | -0.0283 | -0.0283 | -0.0284 | -0.0284 | -0.0285 | -0.0285 | -0.0285 | -0.0285 | -0.0285 | -0.0285 |
| 0.70 microns | -0.0179 | -0.0179 | -0.0179 | -0.0179 | -0.0179 | -0.0179 | -0.0178 | -0.0178 | -0.0178 | -0.0179 | -0.0180 | -0.0180 | -0.0180 | -0.0180 |
| 0.80 microns | -0.0123 | -0.0123 | -0.0123 | -0.0123 | -0.0123 | -0.0123 | -0.0123 | -0.0124 | -0.0124 | -0.0124 | -0.0123 | -0.0123 | -0.0123 | -0.0123 |
| 0.90 microns | -0.0090 | -0.0089 | -0.0090 | -0.0090 | -0.0090 | -0.0090 | -0.0089 | -0.0089 | -0.0088 | -0.0089 | -0.0090 | -0.0090 | -0.0091 | -0.0091 |
| 1.00 microns | -0.0069 | -0.0070 | -0.0069 | -0.0069 | -0.0069 | -0.0069 | -0.0070 | -0.0070 | -0.0069 | -0.0070 | -0.0070 | -0.0070 | -0.0070 | -0.0070 |
| 1.20 microns | -0.0045 | -0.0047 | -0.0047 | -0.0048 | -0.0048 | -0.0049 | -0.0049 | -0.0050 | -0.0048 | -0.0047 | -0.0046 | -0.0045 | -0.0045 | -0.0045 |
| 1.50 microns | -0.0035 | -0.0035 | -0.0035 | -0.0035 | -0.0035 | -0.0035 | -0.0035 | -0.0035 | -0.0035 | -0.0035 | -0.0035 | -0.0035 | -0.0035 | -0.0035 |
| 2.00 microns | -0.0032 | -0.0032 | -0.0032 | -0.0032 | -0.0032 | -0.0032 | -0.0032 | -0.0032 | -0.0032 | -0.0032 | -0.0032 | -0.0032 | -0.0032 | -0.0032 |
| 2.50 microns | -0.0035 | -0.0034 | -0.0034 | -0.0034 | -0.0034 | -0.0034 | -0.0034 | -0.0034 | -0.0034 | -0.0034 | -0.0034 | -0.0034 | -0.0034 | -0.0034 |
| 3.00 microns | -0.0039 | -0.0039 | -0.0039 | -0.0039 | -0.0039 | -0.0039 | -0.0039 | -0.0039 | -0.0039 | -0.0039 | -0.0039 | -0.0039 | -0.0039 | -0.0039 |
| 3.50 microns | -0.0045 | -0.0045 | -0.0045 | -0.0045 | -0.0045 | -0.0045 | -0.0045 | -0.0045 | -0.0045 | -0.0045 | -0.0045 | -0.0045 | -0.0045 | -0.0045 |
| 4.00 microns | -0.0051 | -0.0051 | -0.0051 | -0.0051 | -0.0051 | -0.0051 | -0.0051 | -0.0051 | -0.0051 | -0.0051 | -0.0051 | -0.0051 | -0.0051 | -0.0051 |
| 4.50 microns | -0.0057 | -0.0057 | -0.0057 | -0.0057 | -0.0057 | -0.0057 | -0.0057 | -0.0057 | -0.0057 | -0.0057 | -0.0057 | -0.0057 | -0.0057 | -0.0057 |
| 5.00 microns | -0.0064 | -0.0064 | -0.0064 | -0.0064 | -0.0064 | -0.0064 | -0.0064 | -0.0064 | -0.0064 | -0.0064 | -0.0063 | -0.0063 | -0.0063 | -0.0063 |
| 5.50 microns | -0.0071 | -0.0070 | -0.0070 | -0.0070 | -0.0070 | -0.0071 | -0.0071 | -0.0071 | -0.0071 | -0.0070 | -0.0070 | -0.0070 | -0.0070 | -0.0070 |

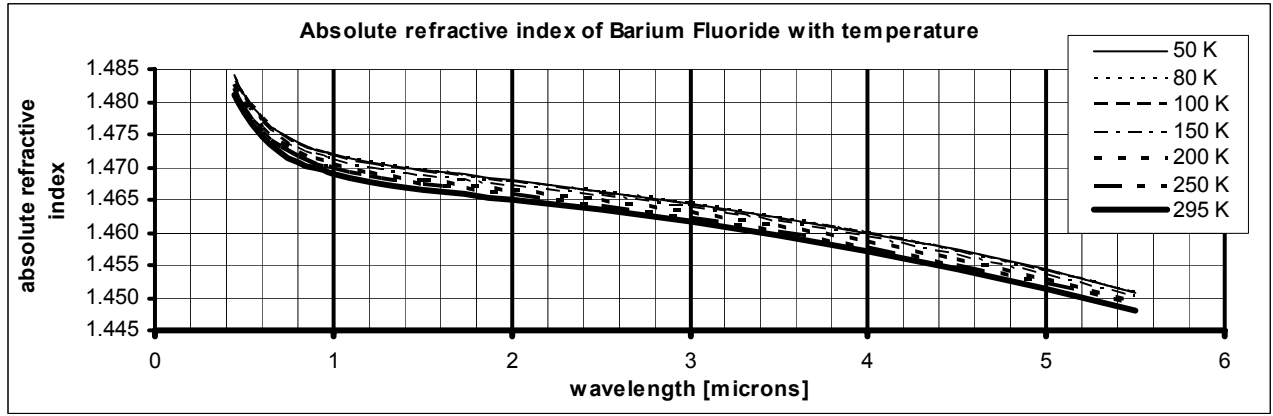

Figure 2 – absolute refractive index (n) for $BaF_2$ at selected wavelengths and temperatures

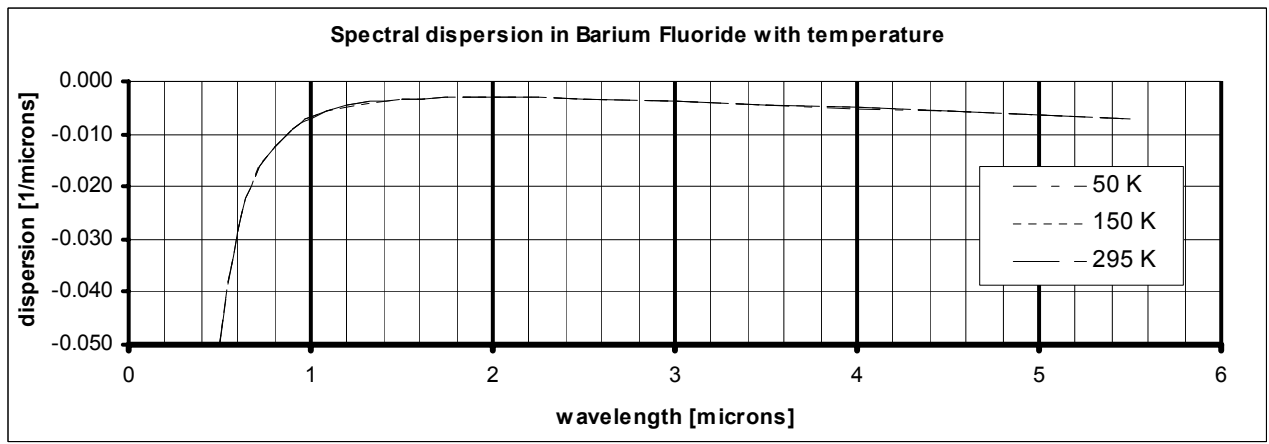

Figure 3 – spectral dispersion in $BaF_2$ at selected temperatures in units of 1/microns

Table 4 – thermo-optic coefficient (dn/dT) of $BaF_2$ at selected wavelengths and temperatures in units of 1/K

| wavelength | 50 K | 60 K | 70 K | 80 K | 90 K | 100 K | 110 K | 120 K | 150 K | 200 K | 250 K | 275 K | 295 K | 300 K |
|---|---|---|---|---|---|---|---|---|---|---|---|---|---|---|
| 0.45 microns | 0.00E+00 | -2.97E-06 | -4.14E-06 | -5.31E-06 | -6.48E-06 | -7.64E-06 | -8.81E-06 | -9.98E-06 | -1.24E-05 | -1.38E-05 | -1.51E-05 | -1.58E-05 | -1.64E-05 | -1.65E-05 |
| 0.50 microns | 0.00E+00 | -3.27E-06 | -4.36E-06 | -5.45E-06 | -6.54E-06 | -7.63E-06 | -8.72E-06 | -9.81E-06 | -1.25E-05 | -1.38E-05 | -1.52E-05 | -1.59E-05 | -1.64E-05 | -1.65E-05 |
| 0.60 microns | 0.00E+00 | -2.80E-06 | -4.06E-06 | -5.32E-06 | -6.58E-06 | -7.84E-06 | -9.10E-06 | -1.04E-05 | -1.27E-05 | -1.40E-05 | -1.52E-05 | -1.59E-05 | -1.64E-05 | -1.65E-05 |
| 0.70 microns | 0.00E+00 | -3.05E-06 | -4.22E-06 | -5.40E-06 | -6.58E-06 | -7.75E-06 | -8.93E-06 | -1.01E-05 | -1.28E-05 | -1.40E-05 | -1.53E-05 | -1.60E-05 | -1.65E-05 | -1.66E-05 |
| 0.80 microns | 0.00E+00 | -2.99E-06 | -4.19E-06 | -5.38E-06 | -6.58E-06 | -7.77E-06 | -8.96E-06 | -1.02E-05 | -1.28E-05 | -1.41E-05 | -1.54E-05 | -1.60E-05 | -1.66E-05 | -1.67E-05 |
| 0.90 microns | 0.00E+00 | -3.05E-06 | -4.23E-06 | -5.42E-06 | -6.60E-06 | -7.78E-06 | -8.97E-06 | -1.02E-05 | -1.27E-05 | -1.41E-05 | -1.54E-05 | -1.61E-05 | -1.66E-05 | -1.67E-05 |
| 1.00 microns | 0.00E+00 | -3.37E-06 | -4.46E-06 | -5.54E-06 | -6.62E-06 | -7.71E-06 | -8.79E-06 | -9.87E-06 | -1.28E-05 | -1.41E-05 | -1.54E-05 | -1.61E-05 | -1.66E-05 | -1.67E-05 |
| 1.20 microns | 0.00E+00 | -3.77E-06 | -4.82E-06 | -5.87E-06 | -6.92E-06 | -7.98E-06 | -9.03E-06 | -1.01E-05 | -1.25E-05 | -1.39E-05 | -1.54E-05 | -1.61E-05 | -1.67E-05 | -1.69E-05 |
| 1.50 microns | 0.00E+00 | -3.92E-06 | -4.94E-06 | -5.96E-06 | -6.98E-06 | -8.00E-06 | -9.02E-06 | -1.00E-05 | -1.24E-05 | -1.39E-05 | -1.54E-05 | -1.62E-05 | -1.68E-05 | -1.70E-05 |
| 2.00 microns | 0.00E+00 | -3.79E-06 | -4.84E-06 | -5.90E-06 | -6.95E-06 | -8.01E-06 | -9.06E-06 | -1.01E-05 | -1.26E-05 | -1.39E-05 | -1.53E-05 | -1.60E-05 | -1.65E-05 | -1.67E-05 |
| 2.50 microns | 0.00E+00 | -4.00E-06 | -4.98E-06 | -5.97E-06 | -6.96E-06 | -7.94E-06 | -8.93E-06 | -9.92E-06 | -1.25E-05 | -1.39E-05 | -1.53E-05 | -1.60E-05 | -1.65E-05 | -1.67E-05 |
| 3.00 microns | 0.00E+00 | -3.54E-06 | -4.66E-06 | -5.79E-06 | -6.91E-06 | -8.03E-06 | -9.16E-06 | -1.03E-05 | -1.24E-05 | -1.38E-05 | -1.52E-05 | -1.59E-05 | -1.65E-05 | -1.67E-05 |
| 3.50 microns | 0.00E+00 | -3.70E-06 | -4.75E-06 | -5.80E-06 | -6.85E-06 | -7.90E-06 | -8.95E-06 | -1.00E-05 | -1.23E-05 | -1.37E-05 | -1.51E-05 | -1.58E-05 | -1.64E-05 | -1.65E-05 |
| 4.00 microns | 0.00E+00 | -3.73E-06 | -4.77E-06 | -5.81E-06 | -6.84E-06 | -7.88E-06 | -8.91E-06 | -9.95E-06 | -1.24E-05 | -1.37E-05 | -1.50E-05 | -1.57E-05 | -1.62E-05 | -1.64E-05 |
| 4.50 microns | 0.00E+00 | -3.62E-06 | -4.67E-06 | -5.72E-06 | -6.77E-06 | -7.82E-06 | -8.87E-06 | -9.92E-06 | -1.24E-05 | -1.36E-05 | -1.48E-05 | -1.53E-05 | -1.58E-05 | -1.59E-05 |
| 5.00 microns | 0.00E+00 | -3.85E-06 | -4.77E-06 | -5.69E-06 | -6.61E-06 | -7.52E-06 | -8.44E-06 | -9.36E-06 | -1.22E-05 | -1.35E-05 | -1.47E-05 | -1.54E-05 | -1.59E-05 | -1.60E-05 |
| 5.50 microns | 0.00E+00 | -3.26E-06 | -4.34E-06 | -5.43E-06 | -6.51E-06 | -7.59E-06 | -8.67E-06 | -9.76E-06 | -1.21E-05 | -1.33E-05 | -1.45E-05 | -1.51E-05 | -1.56E-05 | -1.57E-05 |

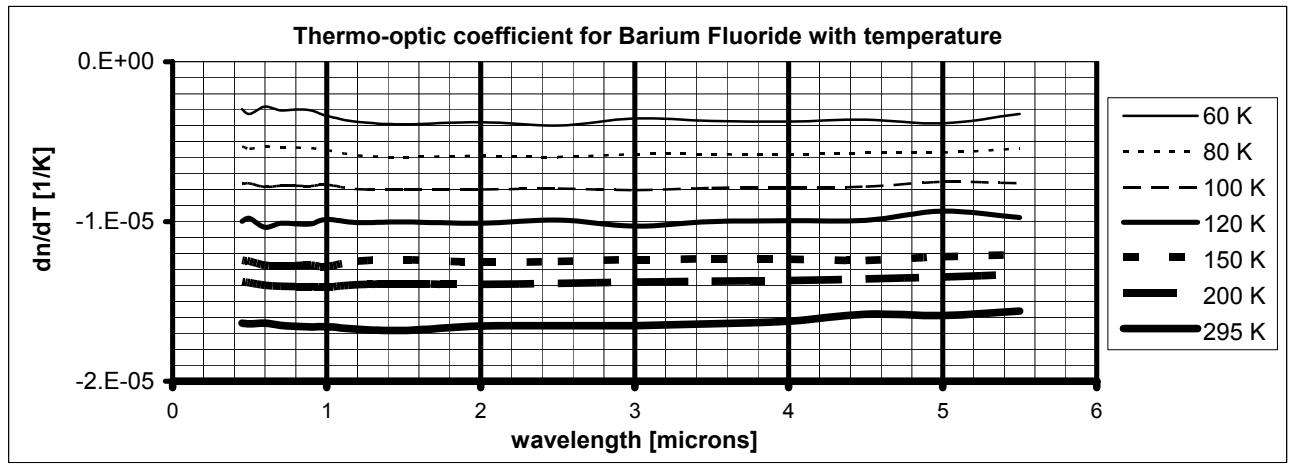

Figure 4 – thermo-optic coefficient (dn/dT) of $BaF_2$ at selected temperatures in units of 1/K

Table 5 – coefficients for the three term Sellmeier model with 4$^{th}$ order temperature dependence for $BaF_2$

| | Coefficients for the temperature dependent Sellmeier equation for $BaF_2$ | | | | | |
| | 50 K <= T <= 300 K; 0.45 μm <= λ <= 5.6 μm | | | | | |
|---|---|---|---|---|---|---|
| | $S_1$ | $S_2$ | $S_3$ | $\lambda_1$ | $\lambda_2$ | $\lambda_3$ |
| Constant term | 8.285359E-01 | 3.315039E-01 | 4.367314E+00 | 8.362026E-02 | -1.148764E-01 | 4.921549E+01 |
| T term | -8.986505E-04 | 9.091254E-04 | -1.161161E-02 | 8.880306E-04 | 3.381142E-03 | -6.672202E-02 |
| $T^2$ term | -1.884197E-06 | 1.656780E-06 | 7.204123E-05 | -1.277585E-05 | -1.897870E-05 | 4.283633E-04 |
| $T^3$ term | -1.332822E-10 | 5.257707E-10 | -4.302326E-08 | 5.231437E-08 | 4.686248E-08 | -3.280396E-07 |
| $T^4$ term | 3.650068E-12 | -3.904140E-12 | -1.764139E-10 | -7.312824E-11 | -4.348650E-11 | -8.848551E-10 |

### 4.2 Lithium Fluoride (LiF)

Absolute refractive indices of LiF were measured over the 0.40 to 5.6 microns wavelength range and over a range of temperatures from 20 to 306 K for two test specimens which, as with $BaF_2$, yielded equal indices to within our measurement uncertainty of +/-2 x 10$^{-5}$. Indices are tabulated in Table 6, plotted in Figure 5 for selected temperatures and wavelengths. Spectral dispersion is tabulated in Table 7, plotted in Figure 6. Thermo-optic coefficient is tabulated in Table 8, plotted in Figure 7. Coefficients for the three term Sellmeier model with 4$^{th}$ order temperature dependence are given in Table 9. We compared our Sellmeier model of LiF at room temperature to the index model of Tropf[6] based on measurements in air by Tilton and Plyler[8] in 1951 multiplied by the index of air. Our model agrees with Tropf's model for LiF at room temperature to within +2.5 x 10$^{-5}$ / -1 x 10$^{-5}$ at all wavelengths.

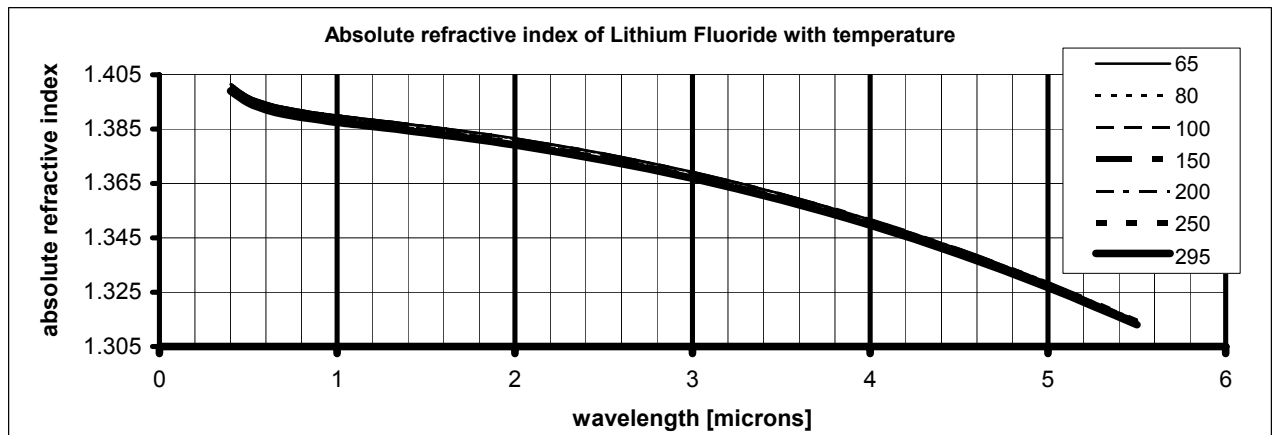

Figure 5 – absolute refractive index of LiF at selected temperatures

Table 6 – absolute refractive index (n) for LiF at selected wavelengths and temperatures

| wavelength | 65 K | 70 K | 75 K | 80 K | 90 K | 100 K | 120 K | 150 K | 200 K | 250 K | 275 K | 295 K | 300 K |
|---|---|---|---|---|---|---|---|---|---|---|---|---|---|
| 0.40 microns | 1.40148 | 1.40147 | 1.40147 | 1.40146 | 1.40143 | 1.40139 | 1.40129 | 1.40105 | 1.40049 | 1.39980 | 1.39941 | 1.39907 | 1.39898 |
| 0.50 microns | 1.39713 | 1.39712 | 1.39711 | 1.39710 | 1.39708 | 1.39704 | 1.39693 | 1.39670 | 1.39611 | 1.39541 | 1.39502 | 1.39468 | 1.39460 |
| 0.60 microns | 1.39467 | 1.39466 | 1.39465 | 1.39464 | 1.39462 | 1.39458 | 1.39447 | 1.39423 | 1.39364 | 1.39293 | 1.39253 | 1.39219 | 1.39210 |
| 0.70 microns | 1.39305 | 1.39304 | 1.39303 | 1.39302 | 1.39300 | 1.39296 | 1.39285 | 1.39261 | 1.39201 | 1.39130 | 1.39090 | 1.39056 | 1.39047 |
| 0.80 microns | 1.39185 | 1.39184 | 1.39184 | 1.39182 | 1.39180 | 1.39176 | 1.39164 | 1.39140 | 1.39081 | 1.39009 | 1.38969 | 1.38935 | 1.38926 |
| 0.90 microns | 1.39087 | 1.39087 | 1.39086 | 1.39085 | 1.39082 | 1.39078 | 1.39067 | 1.39042 | 1.38983 | 1.38911 | 1.38871 | 1.38836 | 1.38827 |
| 1.00 microns | 1.39001 | 1.39000 | 1.39000 | 1.38999 | 1.38996 | 1.38992 | 1.38981 | 1.38957 | 1.38897 | 1.38825 | 1.38785 | 1.38750 | 1.38742 |
| 1.20 microns | 1.38845 | 1.38844 | 1.38842 | 1.38841 | 1.38837 | 1.38833 | 1.38821 | 1.38795 | 1.38736 | 1.38666 | 1.38628 | 1.38595 | 1.38586 |
| 1.50 microns | 1.38608 | 1.38607 | 1.38606 | 1.38604 | 1.38600 | 1.38596 | 1.38584 | 1.38559 | 1.38500 | 1.38431 | 1.38392 | 1.38360 | 1.38351 |
| 2.00 microns | 1.38155 | 1.38154 | 1.38153 | 1.38151 | 1.38148 | 1.38143 | 1.38131 | 1.38108 | 1.38050 | 1.37982 | 1.37944 | 1.37912 | 1.37903 |
| 2.50 microns | 1.37596 | 1.37595 | 1.37593 | 1.37592 | 1.37589 | 1.37584 | 1.37573 | 1.37550 | 1.37494 | 1.37428 | 1.37391 | 1.37360 | 1.37351 |
| 3.00 microns | 1.36915 | 1.36914 | 1.36913 | 1.36912 | 1.36909 | 1.36904 | 1.36893 | 1.36870 | 1.36818 | 1.36754 | 1.36719 | 1.36689 | 1.36681 |
| 3.50 microns | 1.36109 | 1.36108 | 1.36107 | 1.36106 | 1.36103 | 1.36099 | 1.36088 | 1.36067 | 1.36016 | 1.35956 | 1.35922 | 1.35893 | 1.35886 |
| 4.00 microns | 1.35166 | 1.35165 | 1.35164 | 1.35163 | 1.35160 | 1.35157 | 1.35147 | 1.35127 | 1.35079 | 1.35022 | 1.34991 | 1.34964 | 1.34957 |
| 4.50 microns | 1.34083 | 1.34083 | 1.34082 | 1.34080 | 1.34078 | 1.34074 | 1.34065 | 1.34047 | 1.34002 | 1.33950 | 1.33920 | 1.33896 | 1.33889 |
| 5.00 microns | 1.32851 | 1.32850 | 1.32850 | 1.32849 | 1.32847 | 1.32844 | 1.32836 | 1.32818 | 1.32777 | 1.32729 | 1.32702 | 1.32680 | 1.32674 |
| 5.50 microns | 1.31461 | 1.31461 | 1.31461 | 1.31460 | 1.31458 | 1.31455 | 1.31447 | 1.31431 | 1.31392 | 1.31349 | 1.31326 | 1.31308 | 1.31303 |

Table 7 – spectral dispersion (dn/dλ) in LiF at selected wavelengths and temperatures in units of 1/microns

| wavelength | 65 K | 70 K | 75 K | 80 K | 90 K | 100 K | 120 K | 150 K | 200 K | 250 K | 275 K | 295 K | 300 K |
|---|---|---|---|---|---|---|---|---|---|---|---|---|---|
| 0.50 microns | -0.03202 | -0.03201 | -0.03201 | -0.03200 | -0.03200 | -0.03201 | -0.03205 | -0.03217 | -0.03220 | -0.03230 | -0.03238 | -0.03247 | -0.03249 |
| 0.60 microns | -0.01963 | -0.01964 | -0.01964 | -0.01965 | -0.01966 | -0.01967 | -0.01967 | -0.01966 | -0.01973 | -0.01980 | -0.01984 | -0.01988 | -0.01989 |
| 0.70 microns | -0.01373 | -0.01373 | -0.01372 | -0.01372 | -0.01372 | -0.01372 | -0.01372 | -0.01376 | -0.01379 | -0.01382 | -0.01383 | -0.01384 | -0.01384 |
| 0.80 microns | -0.01069 | -0.01070 | -0.01071 | -0.01071 | -0.01072 | -0.01073 | -0.01074 | -0.01074 | -0.01073 | -0.01076 | -0.01078 | -0.01080 | -0.01081 |
| 0.90 microns | -0.00913 | -0.00912 | -0.00911 | -0.00910 | -0.00908 | -0.00907 | -0.00907 | -0.00913 | -0.00910 | -0.00910 | -0.00911 | -0.00913 | -0.00914 |
| 1.00 microns | -0.00822 | -0.00823 | -0.00824 | -0.00825 | -0.00826 | -0.00826 | -0.00825 | -0.00820 | -0.00828 | -0.00828 | -0.00826 | -0.00823 | -0.00822 |
| 1.20 microns | -0.00761 | -0.00765 | -0.00769 | -0.00773 | -0.00780 | -0.00786 | -0.00796 | -0.00806 | -0.00798 | -0.00779 | -0.00766 | -0.00753 | -0.00750 |
| 1.50 microns | -0.00821 | -0.00821 | -0.00820 | -0.00820 | -0.00819 | -0.00818 | -0.00817 | -0.00818 | -0.00814 | -0.00812 | -0.00812 | -0.00812 | -0.00812 |
| 2.00 microns | -0.01007 | -0.01007 | -0.01006 | -0.01006 | -0.01005 | -0.01005 | -0.01004 | -0.01004 | -0.01001 | -0.00998 | -0.00996 | -0.00994 | -0.00994 |
| 2.50 microns | -0.01235 | -0.01235 | -0.01236 | -0.01236 | -0.01236 | -0.01236 | -0.01235 | -0.01230 | -0.01230 | -0.01225 | -0.01223 | -0.01220 | -0.01219 |
| 3.00 microns | -0.01484 | -0.01484 | -0.01484 | -0.01483 | -0.01483 | -0.01482 | -0.01481 | -0.01479 | -0.01475 | -0.01469 | -0.01465 | -0.01462 | -0.01461 |
| 3.50 microns | -0.01749 | -0.01748 | -0.01748 | -0.01747 | -0.01746 | -0.01746 | -0.01745 | -0.01744 | -0.01738 | -0.01731 | -0.01727 | -0.01724 | -0.01723 |
| 4.00 microns | -0.02025 | -0.02025 | -0.02025 | -0.02025 | -0.02025 | -0.02024 | -0.02024 | -0.02020 | -0.02014 | -0.02006 | -0.02001 | -0.01996 | -0.01995 |
| 4.50 microns | -0.02311 | -0.02311 | -0.02311 | -0.02311 | -0.02311 | -0.02311 | -0.02309 | -0.02307 | -0.02300 | -0.02291 | -0.02286 | -0.02281 | -0.02279 |
| 5.00 microns | -0.02619 | -0.02619 | -0.02620 | -0.02620 | -0.02620 | -0.02620 | -0.02618 | -0.02612 | -0.02607 | -0.02598 | -0.02592 | -0.02586 | -0.02584 |
| 5.50 microns | -0.02944 | -0.02946 | -0.02948 | -0.02950 | -0.02952 | -0.02954 | -0.02956 | -0.02949 | -0.02930 | -0.02918 | -0.02914 | -0.02912 | -0.02912 |

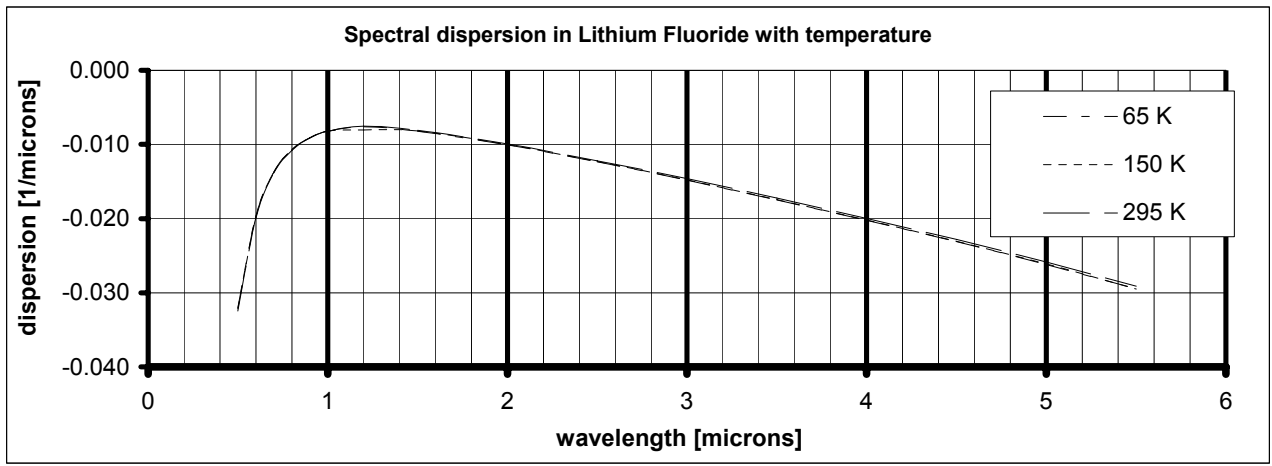

Figure 6 – spectral dispersion in LiF at selected temperatures in units of 1/microns

Table 8 – thermo-optic coefficient (dn/dT) of LiF at selected wavelengths and temperatures in units of 1/K

| wavelength | 65 K | 70 K | 75 K | 80 K | 90 K | 100 K | 120 K | 150 K | 200 K | 250 K | 275 K | 295 K | 300 K |
|---|---|---|---|---|---|---|---|---|---|---|---|---|---|
| 0.40 microns | -7.49E-07 | -1.25E-06 | -1.75E-06 | -2.25E-06 | -3.26E-06 | -4.26E-06 | -6.27E-06 | -9.61E-06 | -1.25E-05 | -1.51E-05 | -1.64E-05 | -1.75E-05 | -1.77E-05 |
| 0.50 microns | -7.86E-07 | -1.30E-06 | -1.82E-06 | -2.33E-06 | -3.36E-06 | -4.40E-06 | -6.46E-06 | -1.02E-05 | -1.30E-05 | -1.51E-05 | -1.62E-05 | -1.71E-05 | -1.73E-05 |
| 0.60 microns | -7.41E-07 | -1.26E-06 | -1.78E-06 | -2.31E-06 | -3.35E-06 | -4.39E-06 | -6.48E-06 | -1.03E-05 | -1.31E-05 | -1.53E-05 | -1.64E-05 | -1.73E-05 | -1.75E-05 |
| 0.70 microns | -6.83E-07 | -1.22E-06 | -1.76E-06 | -2.30E-06 | -3.37E-06 | -4.45E-06 | -6.60E-06 | -1.04E-05 | -1.32E-05 | -1.54E-05 | -1.66E-05 | -1.75E-05 | -1.77E-05 |
| 0.80 microns | -6.53E-07 | -1.20E-06 | -1.75E-06 | -2.31E-06 | -3.41E-06 | -4.51E-06 | -6.71E-06 | -1.05E-05 | -1.32E-05 | -1.55E-05 | -1.66E-05 | -1.75E-05 | -1.78E-05 |
| 0.90 microns | -5.19E-07 | -1.09E-06 | -1.66E-06 | -2.23E-06 | -3.37E-06 | -4.51E-06 | -6.80E-06 | -1.05E-05 | -1.32E-05 | -1.56E-05 | -1.68E-05 | -1.77E-05 | -1.80E-05 |
| 1.00 microns | -8.20E-07 | -1.34E-06 | -1.86E-06 | -2.38E-06 | -3.42E-06 | -4.46E-06 | -6.54E-06 | -1.04E-05 | -1.33E-05 | -1.55E-05 | -1.66E-05 | -1.75E-05 | -1.77E-05 |
| 1.20 microns | -1.79E-06 | -2.26E-06 | -2.73E-06 | -3.21E-06 | -4.15E-06 | -5.10E-06 | -6.99E-06 | -1.03E-05 | -1.29E-05 | -1.50E-05 | -1.60E-05 | -1.69E-05 | -1.71E-05 |
| 1.50 microns | -1.96E-06 | -2.41E-06 | -2.86E-06 | -3.32E-06 | -4.22E-06 | -5.13E-06 | -6.94E-06 | -1.03E-05 | -1.29E-05 | -1.49E-05 | -1.59E-05 | -1.67E-05 | -1.69E-05 |
| 2.00 microns | -1.89E-06 | -2.33E-06 | -2.77E-06 | -3.21E-06 | -4.08E-06 | -4.96E-06 | -6.72E-06 | -1.00E-05 | -1.27E-05 | -1.47E-05 | -1.57E-05 | -1.65E-05 | -1.67E-05 |
| 2.50 microns | -1.80E-06 | -2.23E-06 | -2.65E-06 | -3.08E-06 | -3.93E-06 | -4.79E-06 | -6.50E-06 | -9.70E-06 | -1.23E-05 | -1.43E-05 | -1.52E-05 | -1.60E-05 | -1.62E-05 |
| 3.00 microns | -1.59E-06 | -2.03E-06 | -2.47E-06 | -2.92E-06 | -3.80E-06 | -4.68E-06 | -6.45E-06 | -9.38E-06 | -1.17E-05 | -1.37E-05 | -1.47E-05 | -1.55E-05 | -1.57E-05 |
| 3.50 microns | -1.45E-06 | -1.86E-06 | -2.27E-06 | -2.68E-06 | -3.50E-06 | -4.32E-06 | -5.96E-06 | -8.85E-06 | -1.12E-05 | -1.31E-05 | -1.40E-05 | -1.48E-05 | -1.49E-05 |
| 4.00 microns | -1.34E-06 | -1.73E-06 | -2.12E-06 | -2.52E-06 | -3.30E-06 | -4.09E-06 | -5.66E-06 | -8.38E-06 | -1.05E-05 | -1.23E-05 | -1.31E-05 | -1.38E-05 | -1.40E-05 |
| 4.50 microns | -1.51E-06 | -1.85E-06 | -2.18E-06 | -2.52E-06 | -3.19E-06 | -3.86E-06 | -5.20E-06 | -7.70E-06 | -9.74E-06 | -1.13E-05 | -1.21E-05 | -1.27E-05 | -1.28E-05 |
| 5.00 microns | -7.76E-07 | -1.14E-06 | -1.51E-06 | -1.88E-06 | -2.61E-06 | -3.35E-06 | -4.81E-06 | -7.35E-06 | -8.95E-06 | -1.02E-05 | -1.08E-05 | -1.14E-05 | -1.15E-05 |
| 5.50 microns | -3.12E-07 | -7.13E-07 | -1.11E-06 | -1.51E-06 | -2.32E-06 | -3.12E-06 | -4.72E-06 | -7.46E-06 | -8.32E-06 | -8.85E-06 | -9.11E-06 | -9.32E-06 | -9.38E-06 |

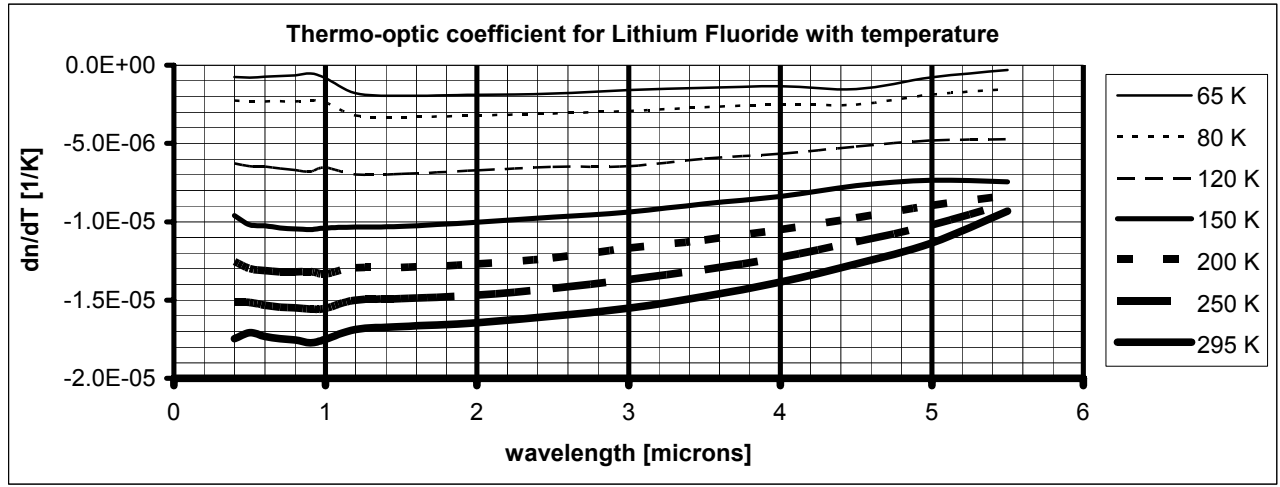

Figure 7 – thermo-optic coefficient (dn/dT) of LiF at selected temperatures in units of 1/K

Table 9 – coefficients for the three term Sellmeier model with 4th order temperature dependence for LiF

| Coefficients for the temperature dependent Sellmeier equation for LiF | | | | | | |
|---|---|---|---|---|---|---|
| 65 K ≤ T ≤ 300 K;  0.40 μm ≤ λ ≤ 5.6 μm | | | | | | |
| | $S_1$ | $S_2$ | $S_3$ | $\lambda_1$ | $\lambda_2$ | $\lambda_3$ |
| Constant term | 9.33385E-01 | 2.91928E+00 | 2.73548E+00 | 7.15768E-02 | 2.93086E+01 | 2.94083E+01 |
| T term | 1.84043E-05 | 6.91076E-03 | 6.71841E-04 | 8.52651E-06 | 3.64286E-02 | 1.04614E-04 |
| $T^2$ term | -1.68868E-07 | 8.47693E-06 | -1.24327E-05 | -5.56075E-08 | 3.22167E-05 | -8.56271E-05 |
| $T^3$ term | 4.08865E-12 | -3.20709E-08 | -2.97405E-08 | 2.37456E-10 | 7.85825E-08 | -1.18786E-07 |
| $T^4$ term | 3.36861E-13 | 1.36856E-10 | 1.21434E-10 | -3.54594E-13 | -2.81432E-10 | 7.82576E-10 |

### 4.3  Zinc Selenide (ZnSe)

Absolute refractive indices of ZnSe were measured over the 0.53 to 5.6 microns wavelength range and over the temperature range from 20 to 310 K for two test specimens which, again, yielded equal indices to within our measurement uncertainty of +/-1 x $10^{-4}$. Indices are tabulated in Table 10, plotted in Figure 8 for selected temperatures and wavelengths.  Spectral dispersion is tabulated in Table 11, plotted in Figure 9.  Thermo-optic coefficient is tabulated in Table 12, plotted in Figure 10. Coefficients for the three term Sellmeier model with 4th order temperature dependence are given in Table 13. We compared our Sellmeier model of ZnSe at room temperature to the index model of Tropf[6] based on measurements in air by Feldman, et al.[9] in 1979 multiplied by the index of air. Our model agrees with Tropf's model for ZnSe at room temperature to within $+3 \times 10^{-5}$ / $-5 \times 10^{-5}$ at all wavelengths. Estimated uncertainty of the measurements used in the Tropf model was $+/-4 \times 10^{-4}$.

Table 10 – absolute refractive index (n) for ZnSe at selected wavelengths and temperatures

| wavelength | 20 K | 25 K | 30 K | 40 K | 60 K | 80 K | 100 K | 120 K | 150 K | 200 K | 250 K | 275 K | 295 K | 300 K |
|---|---|---|---|---|---|---|---|---|---|---|---|---|---|---|
| 0.55 microns | 2.63460 | 2.63472 | 2.63487 | 2.63524 | 2.63630 | 2.63779 | 2.63953 | 2.64166 | 2.64500 | 2.65096 | 2.65742 | 2.66084 | 2.66366 | 2.66438 |
| 0.60 microns | 2.59045 | 2.59056 | 2.59068 | 2.59100 | 2.59193 | 2.59323 | 2.59474 | 2.59655 | 2.59937 | 2.60436 | 2.60973 | 2.61255 | 2.61488 | 2.61547 |
| 0.70 microns | 2.53739 | 2.53749 | 2.53760 | 2.53790 | 2.53870 | 2.53980 | 2.54107 | 2.54258 | 2.54492 | 2.54904 | 2.55343 | 2.55572 | 2.55761 | 2.55809 |
| 0.80 microns | 2.50673 | 2.50682 | 2.50693 | 2.50720 | 2.50794 | 2.50896 | 2.51013 | 2.51149 | 2.51359 | 2.51729 | 2.52122 | 2.52328 | 2.52496 | 2.52539 |
| 0.90 microns | 2.48708 | 2.48716 | 2.48727 | 2.48752 | 2.48822 | 2.48918 | 2.49027 | 2.49155 | 2.49353 | 2.49699 | 2.50065 | 2.50256 | 2.50413 | 2.50452 |
| 1.00 microns | 2.47360 | 2.47368 | 2.47378 | 2.47402 | 2.47469 | 2.47563 | 2.47668 | 2.47790 | 2.47979 | 2.48310 | 2.48659 | 2.48841 | 2.48990 | 2.49028 |
| 1.20 microns | 2.45663 | 2.45671 | 2.45680 | 2.45703 | 2.45766 | 2.45853 | 2.45956 | 2.46072 | 2.46251 | 2.46564 | 2.46895 | 2.47068 | 2.47209 | 2.47244 |
| 1.50 microns | 2.44313 | 2.44321 | 2.44331 | 2.44353 | 2.44415 | 2.44497 | 2.44597 | 2.44708 | 2.44879 | 2.45178 | 2.45495 | 2.45660 | 2.45795 | 2.45829 |
| 2.00 microns | 2.43264 | 2.43272 | 2.43281 | 2.43304 | 2.43364 | 2.43445 | 2.43541 | 2.43648 | 2.43813 | 2.44102 | 2.44408 | 2.44567 | 2.44697 | 2.44730 |
| 2.50 microns | 2.42748 | 2.42756 | 2.42765 | 2.42787 | 2.42846 | 2.42927 | 2.43022 | 2.43127 | 2.43289 | 2.43573 | 2.43874 | 2.44031 | 2.44159 | 2.44192 |
| 3.00 microns | 2.42431 | 2.42438 | 2.42446 | 2.42468 | 2.42526 | 2.42606 | 2.42701 | 2.42806 | 2.42968 | 2.43250 | 2.43547 | 2.43701 | 2.43828 | 2.43860 |
| 3.50 microns | 2.42198 | 2.42205 | 2.42212 | 2.42232 | 2.42289 | 2.42370 | 2.42466 | 2.42571 | 2.42732 | 2.43012 | 2.43307 | 2.43460 | 2.43586 | 2.43617 |
| 4.00 microns | 2.42005 | 2.42010 | 2.42017 | 2.42036 | 2.42091 | 2.42172 | 2.42269 | 2.42374 | 2.42535 | 2.42815 | 2.43108 | 2.43259 | 2.43383 | 2.43414 |
| 4.50 microns | 2.41814 | 2.41822 | 2.41832 | 2.41855 | 2.41916 | 2.41996 | 2.42088 | 2.42190 | 2.42348 | 2.42626 | 2.42921 | 2.43075 | 2.43201 | 2.43233 |
| 5.00 microns | 2.41642 | 2.41650 | 2.41659 | 2.41680 | 2.41740 | 2.41820 | 2.41914 | 2.42016 | 2.42175 | 2.42453 | 2.42746 | 2.42899 | 2.43024 | 2.43055 |
| 5.50 microns | 2.41472 | 2.41479 | 2.41487 | 2.41509 | 2.41568 | 2.41649 | 2.41741 | 2.41845 | 2.42004 | 2.42281 | 2.42572 | 2.42722 | 2.42844 | 2.42875 |

Table 11 – spectral dispersion (dn/dλ) in ZnSe at selected wavelengths and temperatures in units of 1/microns

| wavelength | 20 K | 25 K | 30 K | 40 K | 60 K | 80 K | 100 K | 120 K | 150 K | 200 K | 250 K | 275 K | 295 K | 300 K |
|---|---|---|---|---|---|---|---|---|---|---|---|---|---|---|
| 0.60 microns | -0.74761 | -0.74784 | -0.74812 | -0.74886 | -0.75100 | -0.75404 | -0.75778 | -0.76283 | -0.77096 | -0.78601 | -0.80292 | -0.81207 | -0.81974 | -0.82170 |
| 0.70 microns | -0.39630 | -0.39632 | -0.39637 | -0.39655 | -0.39729 | -0.39852 | -0.40010 | -0.40217 | -0.40545 | -0.41134 | -0.41778 | -0.42121 | -0.42405 | -0.42478 |
| 0.80 microns | -0.24284 | -0.24294 | -0.24304 | -0.24327 | -0.24383 | -0.24452 | -0.24539 | -0.24642 | -0.24808 | -0.25114 | -0.25457 | -0.25643 | -0.25799 | -0.25839 |
| 0.90 microns | -0.16180 | -0.16182 | -0.16185 | -0.16194 | -0.16222 | -0.16265 | -0.16322 | -0.16386 | -0.16488 | -0.16676 | -0.16885 | -0.16997 | -0.17091 | -0.17115 |
| 1.00 microns | -0.11412 | -0.11417 | -0.11422 | -0.11431 | -0.11453 | -0.11477 | -0.11509 | -0.11557 | -0.11631 | -0.11762 | -0.11903 | -0.11977 | -0.12038 | -0.12054 |
| 1.20 microns | -0.06158 | -0.06159 | -0.06162 | -0.06169 | -0.06195 | -0.06235 | -0.06235 | -0.06252 | -0.06279 | -0.06327 | -0.06379 | -0.06407 | -0.06429 | -0.06435 |
| 1.30 microns | -0.04987 | -0.04985 | -0.04985 | -0.04984 | -0.04989 | -0.05002 | -0.05020 | -0.05040 | -0.05070 | -0.05120 | -0.05171 | -0.05197 | -0.05218 | -0.05223 |
| 1.40 microns | -0.03961 | -0.03961 | -0.03961 | -0.03962 | -0.03966 | -0.03975 | -0.03983 | -0.04001 | -0.04028 | -0.04070 | -0.04110 | -0.04130 | -0.04145 | -0.04149 |
| 1.50 microns | -0.03196 | -0.03197 | -0.03199 | -0.03202 | -0.03210 | -0.03219 | -0.03219 | -0.03232 | -0.03252 | -0.03287 | -0.03324 | -0.03344 | -0.03361 | -0.03365 |
| 2.00 microns | -0.01407 | -0.01408 | -0.01408 | -0.01409 | -0.01411 | -0.01414 | -0.01417 | -0.01422 | -0.01429 | -0.01442 | -0.01456 | -0.01463 | -0.01469 | -0.01470 |
| 2.50 microns | -0.00786 | -0.00785 | -0.00784 | -0.00782 | -0.00781 | -0.00783 | -0.00789 | -0.00789 | -0.00790 | -0.00794 | -0.00801 | -0.00805 | -0.00809 | -0.00810 |
| 3.00 microns | -0.00529 | -0.00531 | -0.00532 | -0.00535 | -0.00538 | -0.00538 | -0.00535 | -0.00536 | -0.00537 | -0.00541 | -0.00547 | -0.00550 | -0.00554 | -0.00554 |
| 3.50 microns | -0.00411 | -0.00413 | -0.00415 | -0.00418 | -0.00421 | -0.00421 | -0.00422 | -0.00422 | -0.00423 | -0.00425 | -0.00428 | -0.00431 | -0.00433 | -0.00433 |
| 4.00 microns | -0.00369 | -0.00371 | -0.00372 | -0.00375 | -0.00379 | -0.00380 | -0.00377 | -0.00377 | -0.00377 | -0.00377 | -0.00377 | -0.00378 | -0.00378 | -0.00379 |
| 4.50 microns | -0.00367 | -0.00364 | -0.00362 | -0.00358 | -0.00354 | -0.00354 | -0.00357 | -0.00359 | -0.00361 | -0.00361 | -0.00356 | -0.00351 | -0.00347 | -0.00346 |
| 5.00 microns | -0.00342 | -0.00342 | -0.00341 | -0.00341 | -0.00344 | -0.00349 | -0.00349 | -0.00344 | -0.00339 | -0.00338 | -0.00347 | -0.00355 | -0.00362 | -0.00365 |
| 5.50 microns | -0.00390 | -0.00373 | -0.00359 | -0.00337 | -0.00320 | -0.00340 | -0.00350 | -0.00382 | -0.00417 | -0.00443 | -0.00426 | -0.00402 | -0.00375 | -0.00368 |

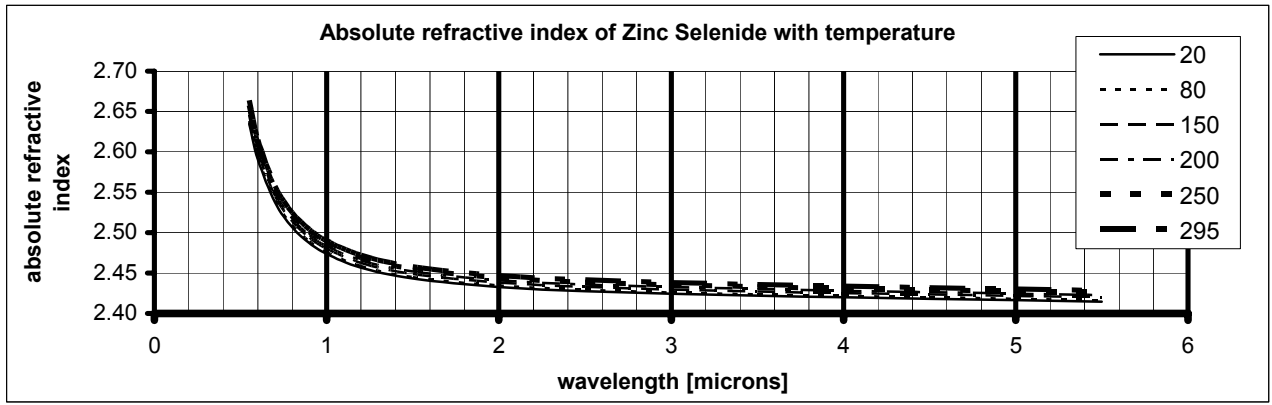

Figure 8 – absolute refractive index of LiF at selected temperatures

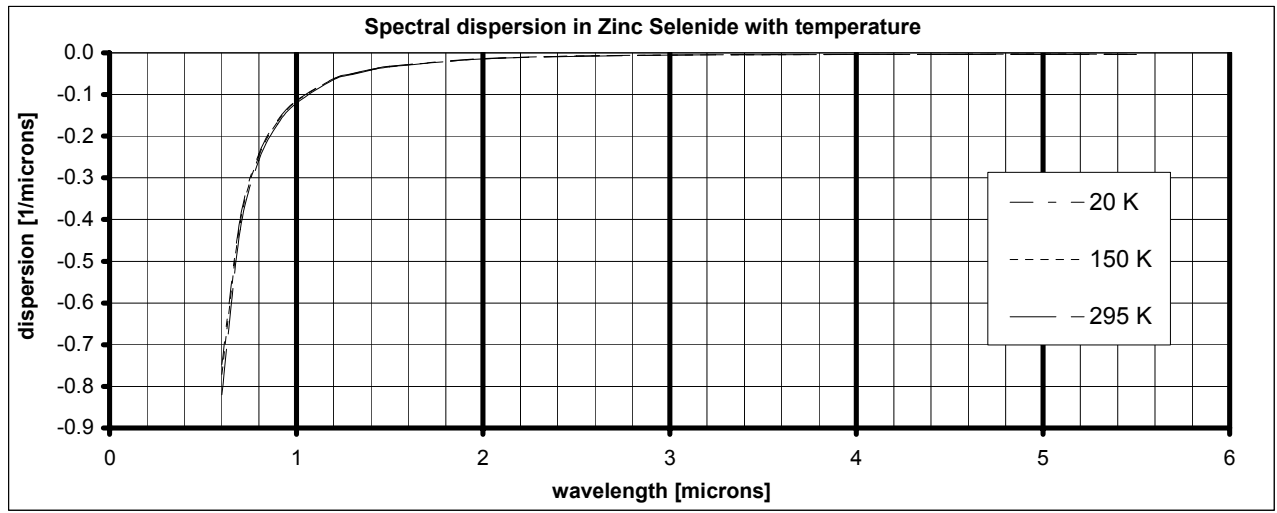

Figure 9 – spectral dispersion in ZnSe at selected temperatures in units of 1/microns

Table 12 – thermo-optic coefficient (dn/dT) of ZnSe at selected wavelengths and temperatures in units of 1/K

| wavelength | 20 K | 25 K | 30 K | 40 K | 60 K | 80 K | 100 K | 120 K | 150 K | 200 K | 250 K | 275 K | 295 K | 300 K |
|---|---|---|---|---|---|---|---|---|---|---|---|---|---|---|
| 0.55 microns | 2.17E-05 | 2.69E-05 | 3.22E-05 | 4.27E-05 | 6.36E-05 | 8.46E-05 | 1.04E-04 | 1.08E-04 | 1.14E-04 | 1.24E-04 | 1.34E-04 | 1.39E-04 | 1.43E-04 | 1.44E-04 |
| 0.60 microns | 1.80E-05 | 2.27E-05 | 2.74E-05 | 3.69E-05 | 5.57E-05 | 7.45E-05 | 8.89E-05 | 9.18E-05 | 9.63E-05 | 1.04E-04 | 1.11E-04 | 1.15E-04 | 1.18E-04 | 1.18E-04 |
| 0.70 microns | 1.82E-05 | 2.19E-05 | 2.55E-05 | 3.29E-05 | 4.76E-05 | 6.23E-05 | 7.43E-05 | 7.64E-05 | 7.97E-05 | 8.51E-05 | 9.05E-05 | 9.32E-05 | 9.53E-05 | 9.59E-05 |
| 0.80 microns | 1.65E-05 | 2.00E-05 | 2.34E-05 | 3.03E-05 | 4.41E-05 | 5.79E-05 | 6.71E-05 | 6.89E-05 | 7.17E-05 | 7.63E-05 | 8.09E-05 | 8.32E-05 | 8.51E-05 | 8.55E-05 |
| 0.90 microns | 1.59E-05 | 1.91E-05 | 2.23E-05 | 2.86E-05 | 4.14E-05 | 5.42E-05 | 6.32E-05 | 6.48E-05 | 6.72E-05 | 7.13E-05 | 7.53E-05 | 7.74E-05 | 7.90E-05 | 7.94E-05 |
| 1.00 microns | 1.47E-05 | 1.79E-05 | 2.11E-05 | 2.75E-05 | 4.02E-05 | 5.29E-05 | 6.04E-05 | 6.19E-05 | 6.42E-05 | 6.80E-05 | 7.18E-05 | 7.37E-05 | 7.52E-05 | 7.56E-05 |
| 1.20 microns | 1.41E-05 | 1.70E-05 | 1.99E-05 | 2.58E-05 | 3.75E-05 | 4.93E-05 | 5.72E-05 | 5.87E-05 | 6.08E-05 | 6.44E-05 | 6.80E-05 | 6.98E-05 | 7.13E-05 | 7.16E-05 |
| 1.50 microns | 1.47E-05 | 1.74E-05 | 2.00E-05 | 2.54E-05 | 3.60E-05 | 4.66E-05 | 5.45E-05 | 5.59E-05 | 5.81E-05 | 6.16E-05 | 6.51E-05 | 6.69E-05 | 6.83E-05 | 6.87E-05 |
| 2.00 microns | 1.43E-05 | 1.69E-05 | 1.96E-05 | 2.48E-05 | 3.54E-05 | 4.60E-05 | 5.27E-05 | 5.41E-05 | 5.61E-05 | 5.95E-05 | 6.29E-05 | 6.45E-05 | 6.59E-05 | 6.62E-05 |
| 2.50 microns | 1.42E-05 | 1.68E-05 | 1.94E-05 | 2.46E-05 | 3.50E-05 | 4.54E-05 | 5.18E-05 | 5.31E-05 | 5.51E-05 | 5.85E-05 | 6.18E-05 | 6.35E-05 | 6.49E-05 | 6.52E-05 |
| 3.00 microns | 1.29E-05 | 1.56E-05 | 1.83E-05 | 2.37E-05 | 3.46E-05 | 4.55E-05 | 5.18E-05 | 5.30E-05 | 5.48E-05 | 5.79E-05 | 6.10E-05 | 6.25E-05 | 6.38E-05 | 6.41E-05 |
| 3.50 microns | 1.13E-05 | 1.42E-05 | 1.70E-05 | 2.28E-05 | 3.43E-05 | 4.59E-05 | 5.16E-05 | 5.28E-05 | 5.46E-05 | 5.76E-05 | 6.05E-05 | 6.20E-05 | 6.32E-05 | 6.35E-05 |
| 4.00 microns | 9.12E-06 | 1.23E-05 | 1.54E-05 | 2.17E-05 | 3.42E-05 | 4.68E-05 | 5.19E-05 | 5.29E-05 | 5.46E-05 | 5.72E-05 | 5.99E-05 | 6.13E-05 | 6.24E-05 | 6.26E-05 |
| 4.50 microns | 1.61E-05 | 1.84E-05 | 2.08E-05 | 2.56E-05 | 3.51E-05 | 4.46E-05 | 5.04E-05 | 5.18E-05 | 5.38E-05 | 5.73E-05 | 6.07E-05 | 6.24E-05 | 6.38E-05 | 6.42E-05 |
| 5.00 microns | 1.36E-05 | 1.63E-05 | 1.90E-05 | 2.43E-05 | 3.50E-05 | 4.58E-05 | 5.08E-05 | 5.20E-05 | 5.39E-05 | 5.71E-05 | 6.02E-05 | 6.18E-05 | 6.31E-05 | 6.34E-05 |
| 5.50 microns | 1.29E-05 | 1.56E-05 | 1.84E-05 | 2.39E-05 | 3.50E-05 | 4.61E-05 | 5.14E-05 | 5.25E-05 | 5.41E-05 | 5.67E-05 | 5.94E-05 | 6.07E-05 | 6.17E-05 | 6.20E-05 |

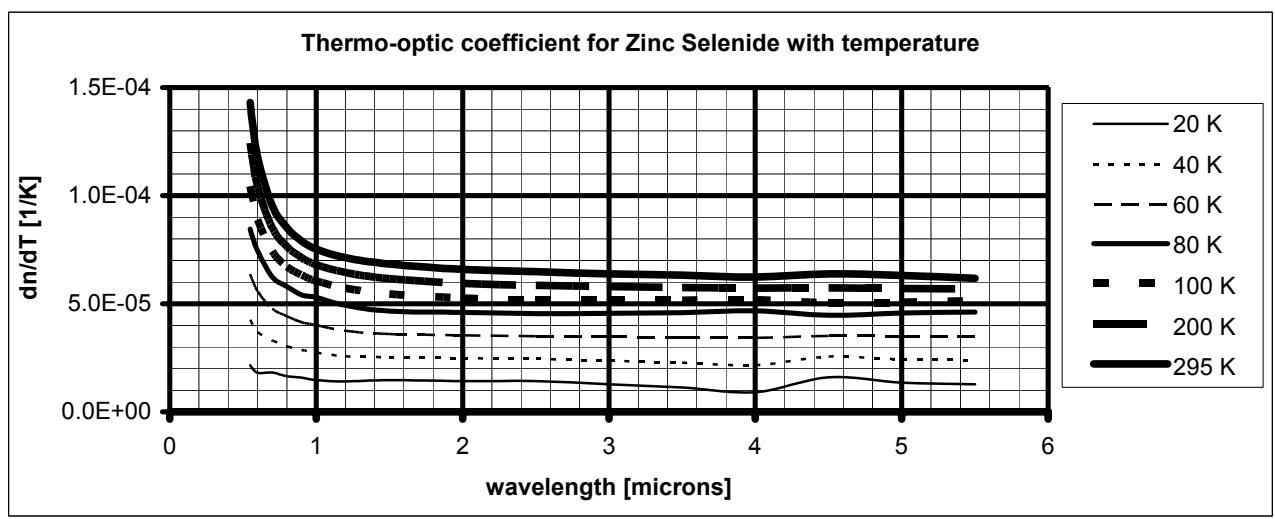

Figure 10 – thermo-optic coefficient (dn/dT) of ZnSe at selected temperatures in units of 1/K

Table 13 – coefficients for the three term Sellmeier model with 4$^{th}$ order temperature dependence for ZnSe

| Coefficients for the temperature dependent Sellmeier equation for ZnSe  20 K ≤ T ≤ 300 K; 0.55 μm ≤ λ ≤ 5.6 μm | | | | | | |
|---|---|---|---|---|---|---|
| | $S_1$ | $S_2$ | $S_3$ | $\lambda_1$ | $\lambda_2$ | $\lambda_3$ |
| **Constant term** | 4.41367E+00 | 4.47774E-01 | 6.70952E+00 | 1.98555E-01 | 3.82382E-01 | 7.33880E+01 |
| **T term** | -1.13389E-03 | 1.11709E-03 | -8.18190E-02 | -3.62359E-05 | -1.56654E-04 | -5.06215E-01 |
| **T$^2$ term** | 2.00829E-05 | -1.80101E-05 | 5.77330E-04 | 7.20678E-07 | 2.56481E-06 | 3.06061E-03 |
| **T$^3$ term** | -8.77087E-08 | 8.10837E-08 | -1.89210E-06 | -3.12380E-09 | -1.07544E-08 | -8.48293E-06 |
| **T$^4$ term** | 1.26557E-10 | -1.18476E-10 | 2.15956E-09 | 4.51629E-12 | 1.53230E-11 | 6.53366E-09 |

## 5. CONCLUSION

Using CHARMS, we have directly measured with unprecedented accuracy the absolute refractive indices of the three infrared materials to be used in the lenses for JWST's NIRCam, from the mid-visible to the mid-infrared, and from room temperature down to as low as 15 K. Good agreement with previous room temperature studies of these materials was demonstrated. The cryogenic temperature at which refractive index stops changing, which we call the saturation temperature, for these materials has been identified. Spectral dispersion appears to be a very weak function of temperature.

Similar measurements of other infrared optical materials, most notably modern synthetic fused silica, are currently underway. Other materials which we will study for NIRCam include Si, ZnS, Ge, and $CaF_2$.

## ACKNOWLEDGEMENTS

The authors wish to thank Dr. Florian Kerber and his colleagues from ESA for their analysis of preliminary index data for ZnSe which was of significant diagnostic value during the accuracy verification process for CHARMS. We also wish to thank Richard Rudy from the Aerospace Corporation for his loaning us several prism samples for evaluation, including the ZnSe sample we used for system diagnostics and for interspecimen comparison purposes.